\documentclass[onecolumn]{emulateapj}
\usepackage{graphicx}
\usepackage{amsmath}
\slugcomment{}
\shortauthors{Kawahara et al.}
\shorttitle{ICM Radial profile and Log-normal Fluctuations}
\newcommand{\rv}{{\boldsymbol{r}}}
\newcommand{\Sigx}{\sigma_{\mathrm{LN}, \, x}}
\newcommand{\SigT}{\sigma_{\mathrm{LN}, \, T}}
\newcommand{\Sign}{\sigma_{\mathrm{LN}, \, n}}
\newcommand{\A}{\langle}
\newcommand{\E}{\rangle}
\newcommand{\Tsl}{T_{\mathrm{sl}}}
\newcommand{\Tew}{T_{\mathrm{ew}}}
\newcommand{\Tmw}{T_{\mathrm{mw}}}
\newcommand{\Tslsm}{T_{\mathrm{sl}}^{\mathrm{sim,m}}}
\newcommand{\Tewsm}{T_{\mathrm{ew}}^{\mathrm{sim,m}}}
\newcommand{\Tslsp}{T_{\mathrm{sl}}^{\mathrm{sim,p}}}
\newcommand{\Tewsp}{T_{\mathrm{ew}}^{\mathrm{sim,p}}}
\newcommand{\Tslm}{T_{\mathrm{sl}}^{\mathrm{model}}}
\newcommand{\Tewm}{T_{\mathrm{ew}}^{\mathrm{model}}}
\newcommand{\Tspec}{T_{\mathrm{spec}}}

\newcommand{\TslP}{T_{\mathrm{sl}}^{\mathrm{RP}}}
\newcommand{\TewP}{T_{\mathrm{ew}}^{\mathrm{RP}}}
\newcommand{\KL}{\kappa^{\mathrm{LI}}}
\newcommand{\KP}{\kappa^{\mathrm{RP}}}
\newcommand{\Kapsim}{\kappa_{\mathrm{sim}}}
\newcommand{\Kapmo}{\kappa_{\mathrm{model}}}
\newcommand{\PLN}{P_\mathrm{LN}}
\newcommand{\PBLN}{P_\mathrm{BLN}}
\newcommand{\Dx}{\delta_x}
\newcommand{\Dn}{\delta_n}
\newcommand{\DT}{\delta_T}

\begin{document}
%
\title{Radial profile and log-normal fluctuations of intra-cluster medium as
an origin of systematic bias of spectroscopic temperature}

\author{
 Hajime Kawahara\altaffilmark{1}, 
 Yasushi Suto\altaffilmark{1}, 
 Tetsu Kitayama\altaffilmark{2}, 
 Shin Sasaki\altaffilmark{3},\\
 Mamoru Shimizu\altaffilmark{1}, 
 Elena Rasia\altaffilmark{4,5,6,7}, 
 and Klaus Dolag\altaffilmark{8} }
\altaffiltext{1}{Department of Physics, The University of Tokyo, 
Tokyo 113-0033, Japan}
\altaffiltext{2}{Department of Physics, Toho University,  Funabashi,
  Chiba 274-8510, Japan}
\altaffiltext{3}{Department of Physics, Tokyo Metropolitan University,
  Hachioji, Tokyo 192-0397, Japan}
\altaffiltext{4}{Dipartimento di Astronomia, 
Universit\`a di Padova, vicolo dell'Osservatorio 2, I-35122 Padova,
Italy}
\altaffiltext{5}{Dipartimento di Astronomia, Universit\`a di Bologna,
ALMA MATER STUDIORUM, via Ranzani 1, 40127, Bologna, Italy}
\altaffiltext{5}{present address: Department of
Physics, 450 Church St., University of Michigan, Ann Arbor, MI
48109-1120}
\altaffiltext{7}{Chandra fellow}
\altaffiltext{8}{Max-Planck Institut fuer Astrophysik,
Karl-Schwarzschild Strasse 1,D-85748 Garching, Germany}
\email{kawahara@utap.phys.s.u-tokyo.ac.jp}
\begin{abstract}
The origin of the recently reported systematic bias in the spectroscopic
temperature of galaxy clusters is investigated using cosmological
hydrodynamical simulations.  We find that the local inhomogeneities of
the gas temperature and density, after corrected for the global radial
profiles, have nearly a universal distribution that resembles the
log-normal function. Based on this log-normal approximation for the
fluctuations in the intra-cluster medium, we develop an analytical model
that explains the bias in the spectroscopic temperature discovered
recently.  We conclude that the multi-phase nature of the intra-cluster
medium not only from the radial profiles but also from the local
inhomogeneities plays an essential role in producing the systematic
bias.
\end{abstract}
\keywords{galaxies: clusters: general -- X-rays:  
galaxies masses -- cosmology: observations}

\section{Introduction}

Recent progress both in numerical simulations and observations has
improved physical modeling of galaxy clusters beyond a simple isothermal
and spherical approximation for a variety of astrophysical and
cosmological applications; departure from isothermal distribution was
discussed in the context of the Sunyaev-Zel'dovich effect
\citep{inagaki95,yoshikawa98,yoshikawa99}, an empirical $\beta$-model
profile has been replaced by that based on the NFW dark matter density
profile \citep{nfw97,makino98,suto98}, non-spherical effects of dark
halos have been considered fairly extensively
\citep{lee98,sheth99,jingsuto02,lee03,lee04,kasun05}, and the physical
model for the origin of the triaxial density profile has been proposed
\citep{lee05}.

Despite an extensive list of the previous studies, no physical
model has been proposed for the statistical nature of underlying
inhomogeneities in the intra-cluster medium (ICM, hereafter). Given the
high spatial resolutions achieved both in observations and simulations,
such a modeling should play a vital role in improving our understanding
of galaxy clusters, which we will attempt to do in this paper.

Temperature of the ICM is one of the most important quantities that
characterize the cluster.  In X-ray observations, the spectroscopic
temperature, $\Tspec$, is estimated by fitting the thermal continuum and
the emission lines of the spectrum. In the presence of inhomogeneities
in the ICM, the temperature so measured is inevitably an {\it averaged}
quantity over a finite sky area and the line-of-sight.  It has been
conventionally assumed that $\Tspec$ is approximately equal to the
emission-weighted temperature:
\begin{equation}
\Tew \equiv \frac{\int n^2 \Lambda (T) T dV}{\int n^2 \Lambda (T) dV},
\label{eq:def_tew}
\end{equation}
where $n$ is the gas number density, $T$ is the gas temperature, and
$\Lambda$ is the cooling function.  \cite{mazzotta04}, however, have
pointed out that $\Tspec$ is systematically lower than $\Tew$. The
authors have proposed an alternative definition for the average, {\it
spectroscopic-like temperature}, as
\begin{equation}
\Tsl \equiv \frac{\int n^2 T^{a-1/2} dV}{\int n^2 T^{a-3/2} dV}.
\label{eq:def_tsl}
\end{equation}
They find that $\Tsl$ with $a=0.75$ reproduces $\Tspec$ within a few
percent for simulated clusters hotter than a few keV, assuming Chandra
or XMM-Newton detector response functions.  \cite{rasia05} performed a
more systematic study of the relation between $\Tew$ and $\Tsl$ using a
sample of clusters from SPH simulations and concluded that $\Tsl \sim
0.7 \Tew$.  \cite{vikhlinin06} provided a useful numerical
routine to compute $\Tsl$ down to ICM temperatures of $\sim 0.5$ keV
with an arbitrary metallicity.  It should be noted that
$\Tew$ is not directly observable, although it is easily obtained from
simulations.

The above bias in the cluster temperature should be properly taken into
account when confronting observational data with theory, 
for example, in cosmological studies. As noted by \cite{rasia05}, it
can result in the offset in the mass-temperature relation of galaxy
clusters. \cite{mshimizu06} have studied its impact on the estimation of
the mass fluctuation amplitude at $8 h^{-1}$Mpc, $\sigma_8$. The authors
perform the statistical analysis using the latest X-ray cluster sample
and find that $\sigma_8 \sim 0.76 \pm 0.01 + 0.50 (1-\alpha_M)$, where
$\alpha_M = \Tspec/\Tew$.  The systematic difference of $\Tspec \sim 0.7
\Tew$ can thus shift $\sigma_8$ by $\sim 0.15$.

In this paper, we aim to  explore the origin of the bias in the
spectroscopic temperature by studying in detail the nature of
inhomogeneities in the ICM.  We investigate both the large-scale
gradient and the small-scale variations of the gas density and
temperature based on the cosmological hydrodynamical simulations. Having
found that the small-scale density and temperature fluctuations
approximately follow the log-normal distributions, we construct an
analytical model for the local ICM inhomogeneities that can
simultaneously explain the systematic bias.

The plan of this paper is as follows. In \S 2, we describe our
simulation data, construct the mock spectra and compare quantitatively
the spectroscopic temperature with the emission-weighted temperature and
the spectroscopic-like temperature suggested by \cite{mazzotta04}.  In
\S 3, we propose an analytical model for the inhomogeneities in the
ICM. In \S 4, we test our model against the result of the
simulation. Finally, we summarize our conclusions in \S 5.  Throughout
the paper, temperatures are measured in units of keV.

\section{The Bias in  the Spectroscopic  Temperature}
  
\subsection{Cosmological Hydrodynamical Simulation}

The results presented in this paper have been obtained by using the
final output of the Smoothing Particle Hydrodynamic (SPH) simulation of
the local universe performed by \cite{dolag05}.  The initial conditions
were similar to those adopted by \citet{Mathis2002} in their study
(based on a pure N-body simulation) of structure formation in the local
universe. The simulation assumes the spatially--flat $\Lambda$ cold dark
matter ($\Lambda$ CDM) universe with a present matter density parameter
$\Omega_{0m}=0.3$, a dimensionless Hubble parameter $h = H_0/100 ~{\rm
km~s^{-1}~Mpc^{-1}} = 0.7$, an rms density fluctuation amplitude
$\sigma_8 = 0.9$ and a baryon density parameter $\Omega_{\rm
b}=0.04$. Both the number of dark matter and SPH particles are $\sim 50$
million within the high-resolution sphere of radius $\sim 110$ Mpc which
is embedded in a periodic box of $\sim 343$ Mpc on a side, filled with
nearly 7 million low-resolution dark matter particles.  The simulation
is designed to reproduce the matter distribution of the local Universe
by adopting the initial conditions based on the {\it IRAS} galaxy
distribution smoothed on a scale of 7 Mpc \citep[see][for detail]
{Mathis2002}.

The run has been carried out with {\small GADGET-2}
\citep{2005MNRAS.364.1105S}, a new version of the parallel Tree-SPH
simulation code {\small GADGET} \citep{SP01.1}.  The code uses an
entropy-conserving formulation of SPH \citep{2002MNRAS.333..649S}, and
allows a treatment of radiative cooling, heating by a UV background, and
star formation and feedback processes.  The latter is based on a
sub-resolution model for the multiphase structure of the interstellar
medium \citep{2003MNRAS.339..289S}; in short, each SPH particle is
assumed to represent a two-phase fluid consisting of cold clouds and
ambient hot gas.

 The code also follows the pattern of metal production from the past
history of cosmic star formation \citep{2004MNRAS.349L..19T}.  This is
done by computing the contributions from both Type-II and Type-Ia
supernovae and energy feedback and metals are released gradually in
time, accordingly to the appropriate lifetimes of the different stellar
populations. This treatment also includes in a self-consistent way the
dependence of the gas cooling on the local metallicity. The feedback
scheme assumes a Salpeter IMF \citep{1955ApJ...121..161S} and its
parameters have been fixed to get a wind velocity of $\approx 480$
$\mathrm{km~s^{-1}}$.  In a typical massive cluster the SNe (II and Ia)
add to the ICM as feedback $\approx 2$ keV per particle in an Hubble
time (assuming a cosmological mixture of H and He); $\approx 25$ per
cent of this energy goes into winds.  A more detailed discussion of
cluster properties and metal distribution within the ICM as resulting in
simulations including the metal enrichment feedback scheme can be found
in \citet{2004MNRAS.349L..19T}. The simulation provides the
metallicities of the six different species for each SPH particle.  Given
the fact that the major question that we addressed is not the accurate
estimate of $\Tspec$ or $\Tsl$, but the systematic difference between
the two, we decided to avoid the unnecessary complication and simply to
assume the constant metallicity.  Therefore we adopt a constant
metallicity of 0.3 $Z_\odot$ in constructing mock spectra below, and the
MEKAL (not VMEKAL) model for the spectral fitting. 

 The gravitational force resolution (i.e. the comoving softening length)
of the simulations has been fixed to be 14 kpc (Plummer-equivalent),
which is comparable to the inter-particle separation reached by the SPH
particles in the dense centers of our simulated galaxy clusters.

\subsection{Mock Spectra of Simulated Clusters}

Among the most massive clusters formed within the simulation we
extracted six mock galaxy clusters, contrived to resemble A3627, Hydra,
Perseus, Virgo, Coma, and Centaurus, respectively.  Table \ref{tab:all}
lists the observed and simulated values of the total mass and the radius
of these clusters.  In order to specify the degree of the bias in our
simulated clusters, we create the mock spectra and compute $\Tspec$ in
the following manner.

First, we extract a $3 h^{-1}~ \mathrm{ Mpc }$ cubic region around the
center of a simulated cluster and divide it into $256^3$ cells so that
the size of each cell is approximately equal to the gravitational
softening length mentioned above.  The center of each cluster is
assigned so that the center of a sphere with radius $1 h^{-1}~
\mathrm{Mpc}$ equals to the center of mass of dark matter and baryon
within the sphere.

 The gas density and temperature of
each mesh point (labeled by $I$) are calculated  using the SPH
particles as
\begin{eqnarray}
\rho_I = \sum_{i=1}^{Ngas} m_i W(|\rv_I-\rv_i|,h_i), \\
T_I = \sum_{i=1}^{Ngas} \frac{m_i T_i}{\rho_i} W(|\rv_I-\rv_i|,h_i), 
\end{eqnarray}
where  $\rv_I$ is the position of the mesh point, $W$
denotes the smoothing kernel, and $m_i$, $\rv_i$, $h_i$, $T_i$, and $\rho_i$
are the mass associated with the hot phase, position, smoothing length,
temperature, and density associated with the hot gas phase of
the $i$-th SPH particle, respectively. We adopt the smoothing
kernel: 
\begin{eqnarray}
     W(|\rv_I-\rv_i|,h_i) = \frac{1}{\pi h_i^3} 
      \left\{
       \begin{array}{lr}
	1 - (3/2) u^2 + (3/4)u^3 & \mbox{if $0 \le u \le 1$} \\
	(2 - u)^3/4 & \mbox{if $1 \le u \le 2$} \\
	0 & \mbox{otherwise,} \\
       \end{array}
      \right. 
\end{eqnarray}
where $u \equiv |\rv_I-\rv_i|/h_i$.

\begin{figure}[htbp]
{\includegraphics[width=0.8 \linewidth]{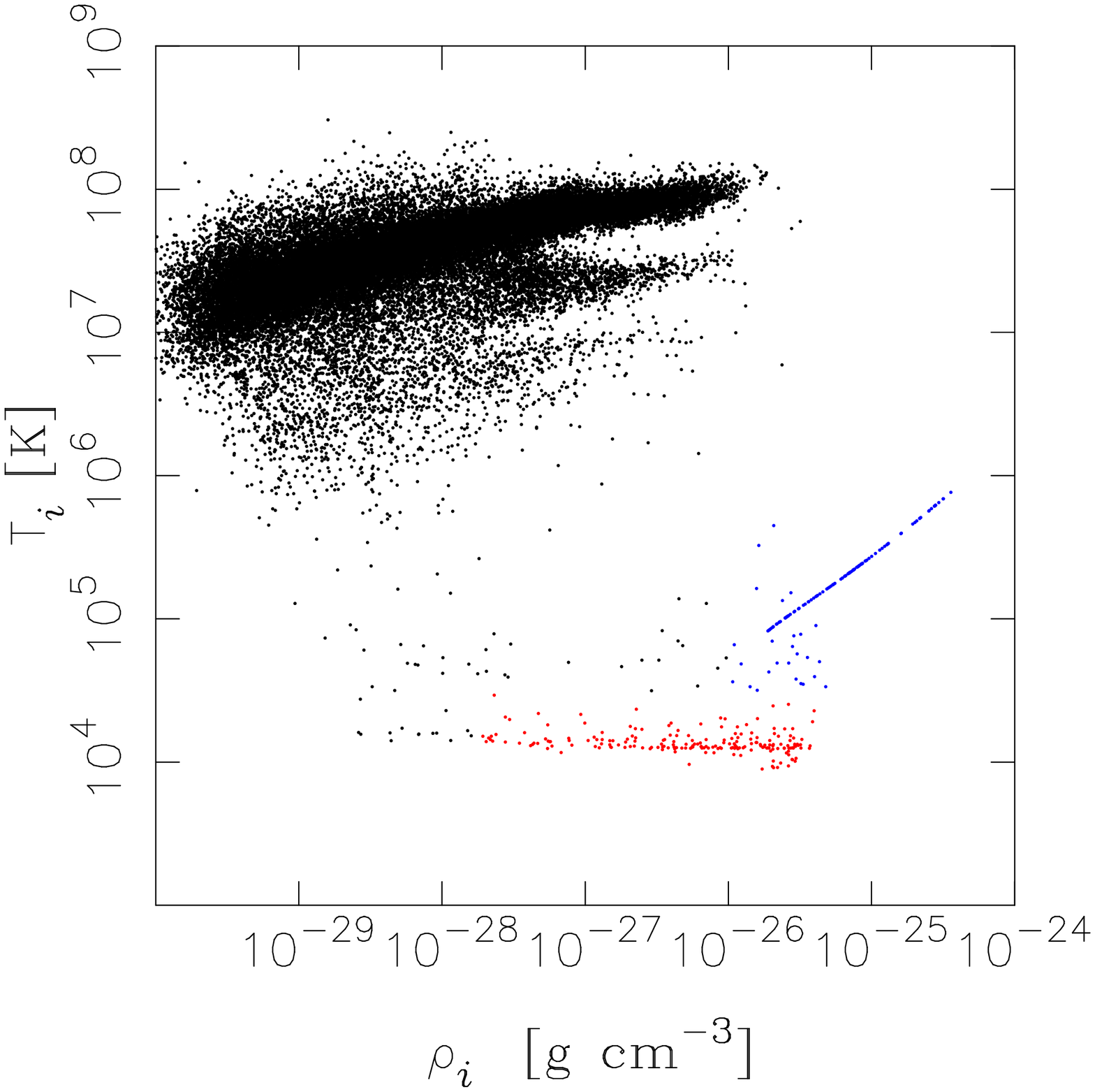}}
 \caption{Scatter plots of temperatures and densities of SPH particles.
Red and blue points indicate particles with unphysical 
temperatures and densities which are removed in computing the X-ray
 emission.
\label{fig:scatter}}
\end{figure}

It should be noted that the current implementation of the SPH simulation
results in a small fraction of SPH particles that have unphysical
temperatures and densities. This is shown in the temperature -- density
scatter plot of Figure \ref{fig:scatter}. The red points correspond to
SPH particles that should be sufficiently cooled, but not here because
of the limited resolution of the simulation.  Thus if they satisfy the
Jean criterion, they should be regarded as simply cold clumps without
retaining the hot gas nature; see Figure 1 and section 2.1 of
\citet{yoshikawa01}.  In contrast, the blue points represent the SPH
particles that have experienced the cooling catastrophe, and have
significantly high cold gas fraction (larger than 10 percent).  In
either case, they are not supposed to contribute the X-ray emission.
Thus we remove their spurious contribution to the X-ray emission and the
temperature estimate of ICM. Specifically we follow \citet{borgani04},
and exclude particles (red points) with $T_i < 3 \times 10^4 \mathrm{K}$
and $\rho_i > 500 \rho_{\rm c} \Omega_{\rm b}$ , where $\rho_{\rm c}$ is
the critical density, and particles (blue points) with more than ten
percent mass fraction of the cold phase.  While the total mass of the
excluded particles is very small ($\sim 1 \%$), they occupy a specific
region in the $\rho$-$T$ plane and leave some spurious signal due to
high density, in particular for blue points.

Second, we compute the photon flux $f(E)$ from the mesh points
within the radius $r_{200}$ from the cluster center as
\begin{eqnarray}
f(E) \, dE \propto \exp{(-\sigma_{gal}(E) N_H)} \sum_{I \in r_{200}} \frac{\rho_I^2}{4 \pi (1+z_{\mathrm{cl}})^4}
		\left( \frac{X}{m_\mathrm{p}^2}\right) \, P_{\mathrm{em}}(T_I,Z,E(1+z_{\mathrm{cl}})) \, dE (1+z_{\mathrm{cl}}),
\end{eqnarray}
where $z_{\mathrm{cl}}$ denotes  the redshift of the
simulated cluster, $Z$ is the metallicity (we adopt 0.3 $Z_{\odot}$),
 $X$ is the hydrogen mass fraction, $m_p$ is the proton
mass and $P_{\mathrm{em}}(T_I,Z,E)$ is the emissivity assuming
collisional ionization equilibrium.  We calculate
$P_{\mathrm{em}}(T_I,Z,E)$ using SPEX 2.0. The term $\exp{(-\sigma_{gal}
N_H)}$ represents the galactic extinction; $N_H$ is the column density
of hydrogen and $\sigma_{gal} (E)$ is the absorption cross section of
\cite{Morrison1983}.  Since we are interested in the effect due to the spectrum
distortion, not statistical error,  we adopt a long exposure time as the total photon
counts ${\cal N} = \int_{E=0.5~\mathrm{keV}}^{E=10~\mathrm{keV}} E f(E)
\, dE \sim$ 500,000.  In this paper, we consider mock
observations using {\it Chandra} and {\it XMM-Newton}, thus we neglect
a peculiar velocity of the cluster and a turbulent velocity in ICM
because of insufficient energy resolution of {\it Chandra} ACIS-S3 and
{\it XMM-Newton} MOS1 detector.

Finally, the mock observed spectra are created by XSPEC version 12.0.
We consider three cases for the detector response corresponding to 1)
perfect response, 2) {\it Chandra} ACIS-S3, and 3) {\it XMM-Newton}
MOS1. In the first case, we also assume no galactic extinction ($N_H=0$)
and refer to it as an ``IDEAL'' case. In the second and third cases, we
adopt an observed value to $N_H$ listed in Table \ref{tab:all} and
redistribute the photon counts of the detector channel according to RMF
(redistribution matrix file) of ACIS-S3 and MOS1 using the {\it
rejection method}.
 
\begin{figure}[htbp]
{\includegraphics[width=0.8 \linewidth]{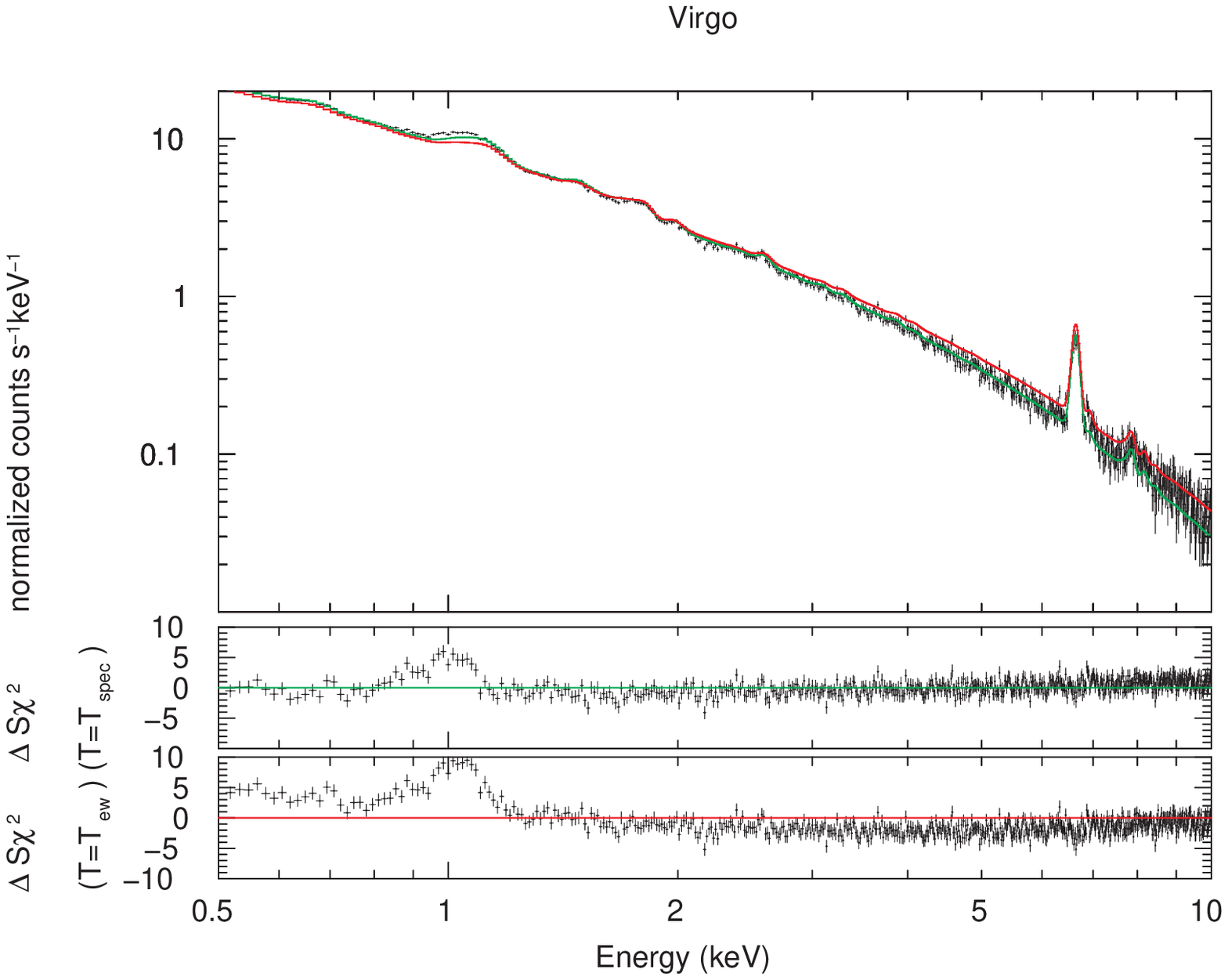}}
{\includegraphics[width=0.8 \linewidth]{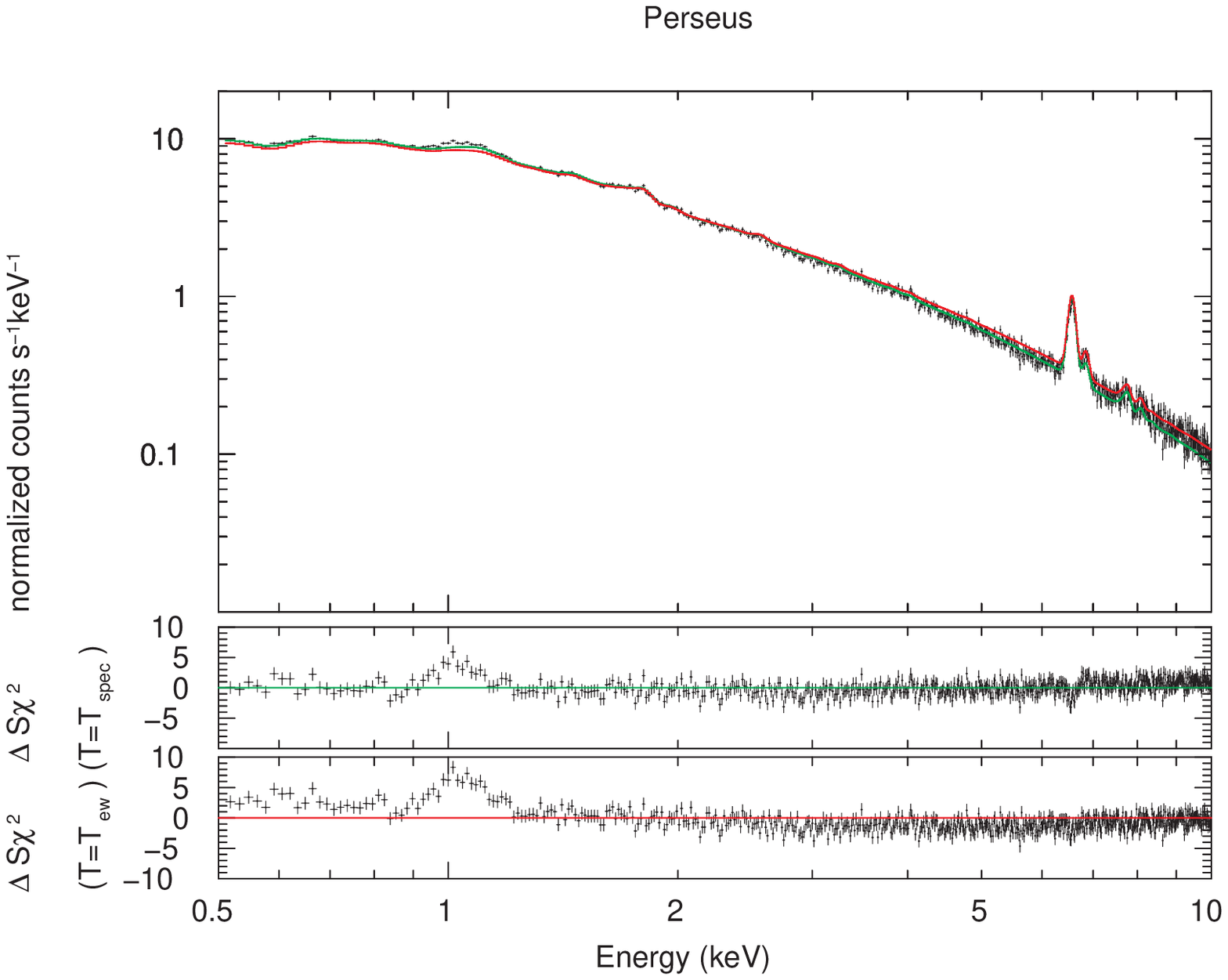}}
 \caption{The examples of the mock spectrum of two simulated
 clusters.  The top panel shows the  results
 for ``Virgo'' and the bottom panel for ``Perseus.''
 Black marks are the mock spectra. Green line
 provides the best-fit spectrum, Red line is that of a single temperature
 thermal model with the temperature $T=\Tewsm$. Each panel
 has two residuals in terms of sigmas with error bar of
 size one. Upper one is the residual of the best-fit spectrum
 ($T=\Tspec$). Lower one is that of the thermal model  with
 the emission weighted} temperature ($T=\Tewsm$). 
  
\label{fig:exam_spec}
\end{figure}

Figure \ref{fig:exam_spec} illustrates the mock spectra of ``Virgo'' and
``Perseus'' using RMF of ACIS-S3. Unless stated otherwise, we fit the
spectra by an absorbed single-temperature MEKAL model in the energy band
0.5-10.0 keV.  We define the spectroscopic temperature, $\Tspec$, as the
best-fit temperature provided by this procedure. Since the spatial
resolution of the current simulations is not sufficient to fully resolve
the cooling central regions, a single-temperature model yields a
reasonable fit to the mock spectra.  For comparison, we also plot the
spectra for a single temperature corresponding to the ``emission
weighted'' value of the mesh points within $r_{200}$:
\begin{equation}
\Tewsm = \frac{\sum_{I \in r_{200}} \rho_I^2 T_I \Lambda(T_I)}
{\sum_{I \in r_{200}}
 \rho_I^2 \Lambda(T_I)}. 
\label{eq:tmesh}
\end{equation}
We calculate the cooling function $\Lambda (T)$ using SPEX 2.0 assuming
collisional ionization equilibrium, the energy range of 0.5-10.0 keV,
and the metallicity $0.3 Z_{\odot}$.  The difference between $\Tspec$
and $\Tewsm$ is clearly distinguishable on the spectral basis in the
current detectors.

\begin{figure}[htbp]
\plotone{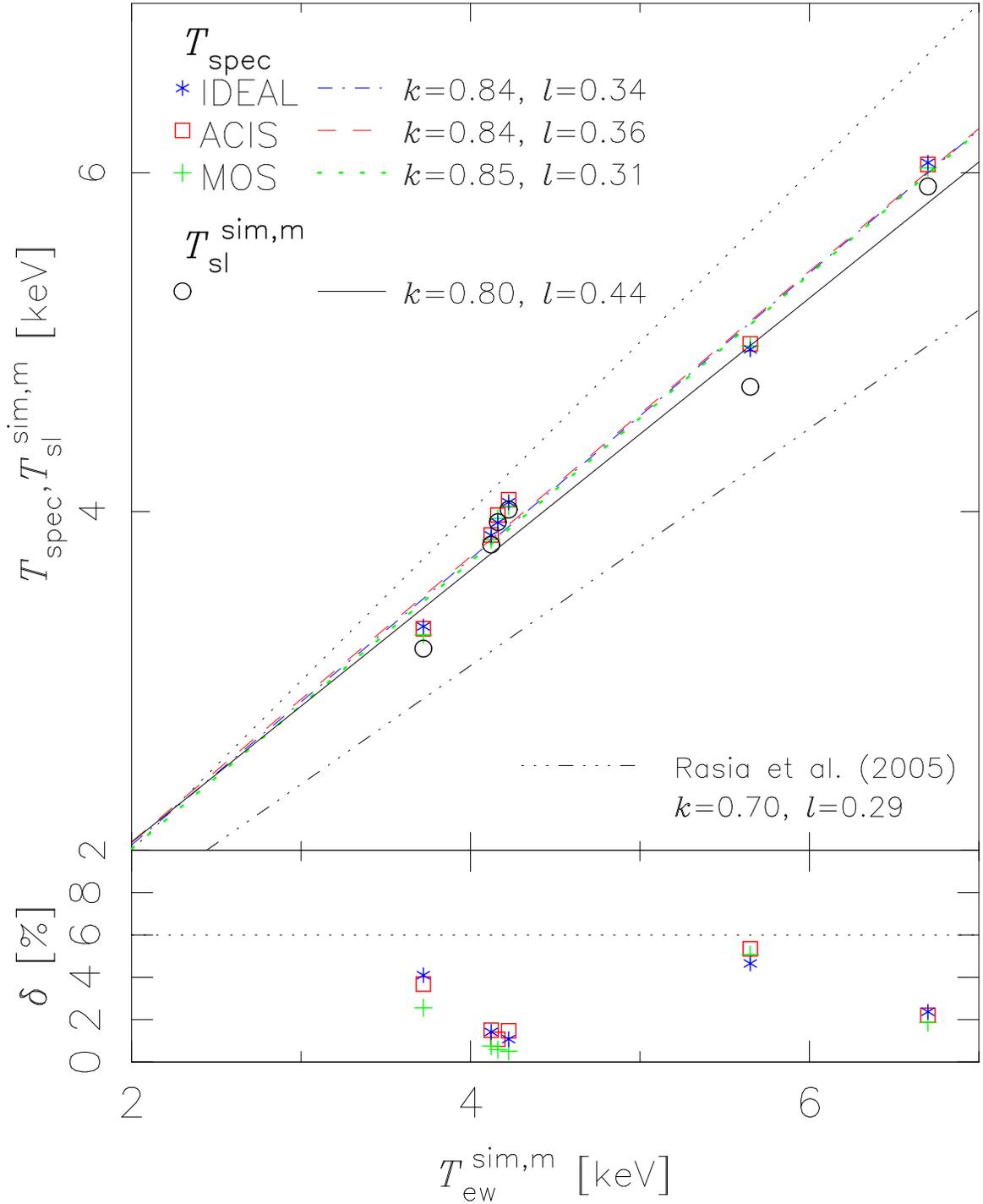}
  \caption{ The upper panel shows the relation of $\Tewsm$
 to $\Tspec$  (asterisk shows IDEAL, 
 square ACIS, and plus MOS) and  $\Tslsm$ (open circle).
 The lower panel shows
the difference between $\Tspec$ and $\Tslsm$: $\delta \equiv 100
 (\Tspec/\Tslsm -1 ) [\%]$, where $\Tspec$ is the best-fit
 temperature of the mock spectra  and $\Tslsm$ is given 
 by equation (\ref{eq:tslmesh}).}
\label{fig:Tsl-Tspec}
\end{figure}

Figure \ref{fig:Tsl-Tspec} shows the relation between $\Tspec$ and
$\Tew$ for our sample of simulated clusters. It is well represented by a
linear relation $\Tspec =k \Tew + l$ with the fitted values of $k=0.84,
l=0.34$ (IDEAL), $k=0.84, l=0.36$ (ACIS) and $k=0.85, l=0.31$ (MOS),
respectively. In the range of temperatures corresponding to rich
clusters, the spectroscopic temperature $\Tspec$ is systematically lower
than $\Tew$ by $10-20$ \%.

We note that the above bias should depend on the energy band in which
$\Tspec$ is evaluated. In order to demonstrate it quantitatively, we
also list in Table \ref{tab:all} the fitted values of $\Tspec$ from the
0.1-2.4 keV and 2.0-10.0 keV data, respectively.  Because
the exponential tail of the thermal bremsstrahlung spectrum from hotter
components has negligible contribution in the softer band, the bias
tends to increase and decrease in the softer and harder bands,
respectively.

\subsection{Spectroscopic-Like Temperature}

In order to better approximate $\Tspec$, \cite{mazzotta04} proposed a ``
spectroscopic-like temperature''; they found that equation
(\ref{eq:def_tsl}) with $a=0.75$ reproduces $\Tspec$ in the 0.5-10.0
keV band within a few percent.  Throughout this paper, we adopt $a=0.75$
when we estimate the spectroscopic-like temperature quantitatively.  In
Figure \ref{fig:Tsl-Tspec}, we also plot this quantity computed from the
mesh points within $r_{200}$:
\begin{equation}
\Tslsm = \frac{\sum_{I \in r_{200}} \rho_I^2 T_I^{0.25}}{\sum_{I \in r_{200}}
 \rho_I^2 T^{-0.75}_I}.
\label{eq:tslmesh}
\end{equation}
As indicated in the bottom panel, $\Tslsm$ reproduces $\Tspec$ within 6
\% for all the simulated clusters in our sample.  Given this agreement,
we hereafter use $\Tslsm$ to represent $\Tspec$, and express the
bias in the spectroscopic  temperature by
\begin{equation}   
\Kapsim \equiv \Tslsm/\Tewsm.
\end{equation}

Table \ref{tab:all} provides $\Kapsim$ (mesh-wise) for the six simulated
clusters. The range of $\Kapsim$ (mesh-wise) is approximately 0.8-0.9.
While $\Kapsim$ is systematically lower than unity, the value is
somewhat higher than the results of \cite{rasia05}, $\Tsl \sim 0.7
\Tew$.  This is likely due to the different physics incorporated in the
simulations and the difference in how $\Tsl$ and $\Tew$ are computed
from the simulation outputs.  The major difference of the
physics is the amplitude of the wind velocity employed; \cite{rasia05}
used have a feedback with weaker wind of $340 \mathrm{km \, s^{-1}}$,
while our current simulations adopt a higher value of $480 \mathrm{km \,
s^{-1}}$. Because weaker wind cannot remove small cold blobs
effectively, the value of $\Tsl/\Tew$ of \cite{rasia05} is
 expected to be larger.

To show the difference of the temperature computation scheme explicitly,
we also list in Table \ref{tab:all} the values of $\Kapsim \equiv
\Tslsp/\Tewsp$ (particle-wise) computed in the ``particle-wise''
definitions used in \cite{rasia05}:
\begin{eqnarray}
\Tewsp = \frac{\sum_{i \in r_{200}} m_{i} \rho_{i} \Lambda(T_i) T_i
 }{\sum_{i \in r_{200}} m_{i} \rho_{i} \Lambda(T_i)},
\label{eq:p-tew}
\end{eqnarray}
and
\begin{eqnarray}
\Tslsp = \frac{\sum_{i \in r_{200}} m_{i} \rho_{i}  T_i^{0.25}
 }{\sum_{i \in r_{200}} m_{i} \rho_{i} T_i^{-0.75}},
\label{eq:p-tsl}
\end{eqnarray}
\citep[see also][]{borgani04}.  In practice, the emission-weighted and
spectroscopic-like temperatures defined in equation (\ref{eq:p-tew}) and
equation (\ref{eq:p-tsl}) are sensitive to a small number of cold (and
dense) SPH particles present, while their contribution is negligible in
the mesh-wise definitions, equation (\ref{eq:tmesh}) and equation
(\ref{eq:tslmesh}). As in \cite{rasia05}, we have removed the SPH
particles below a threshold temperature $T_{lim}=0.5$ keV to compute
$\Kapsim$ (particle-wise).  Table \ref{tab:all} indicates that $\Kapsim$
(particle-wise) tends to be systematically smaller than $\Kapsim$
(mesh-wise). Even adopting $T_{lim}=0.01$ keV makes $\Kapsim$
(particle-wise) smaller only by a few percent.

Given the limit of the particle-wise definitions mentioned above, we use
the mesh-wise definitions of the emission-weighted temperature
($\Tewsm$) and the spectroscopic-like temperature ($\Tslsm$) given in
equation (\ref{eq:tmesh}) and (\ref{eq:tslmesh}), respectively, in the
following sections.

\section{Origin of the Bias in the Spectroscopic  Temperature}

\subsection{Radial Profile and Log-normal Distribution of 
Temperature and Density}
Having quantified the bias in the temperature of simulated clusters, we
investigate its physical origin in greater detail. Since clusters in
general exhibit inhomogeneities over various scales, we begin with
segregating the large-scale gradient and the small-scale fluctuations of
the gas density and temperature.

\begin{figure}[htbp]
 \plotone{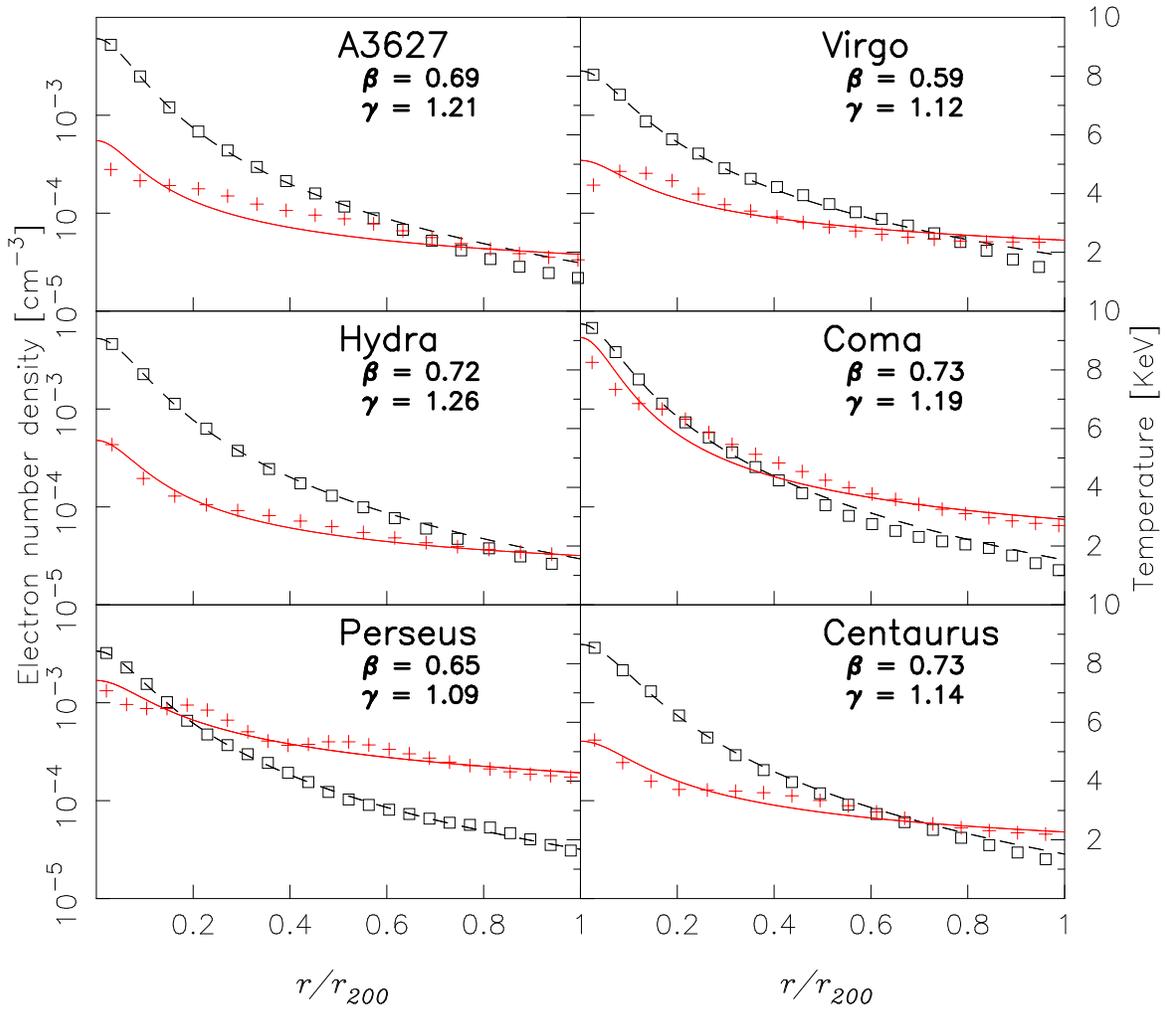}
  \caption{The radial profile of simulated clusters. Square
  provides the (electron number) density profile and dashed line is its
  fitting line assuming the beta model. Plus shows the temperature
  profile and solid line is its fitting line assuming the polytropic
  model. Each square and plus point corresponds to the shell with a
 width of $67 h^{-1}~ \mathrm{kpc}$.}
\label{fig:radp}
\end{figure}

For the large-scale gradient, we use the radially averaged profile of
the gas temperature and density shown in Figure \ref{fig:radp}. We
divide the simulated clusters into spherical shells with a width of
$67 h^{-1}~ \mathrm{ kpc }$ and calculate the average temperature $T(r)$ and
density $n(r)$ in each shell. The density profile $n(r)$ is fitted to
the conventional beta model given by
\begin{equation}
n(r) = n_0 \,\left[ \frac{1}{ 1 + (r/r_c)^2 } \right]^{3 \beta/2},
\label{eq:beta-model}
\end{equation}
where $n_0$ is the central density, $r_c$ is the core radius, and
$\beta$ is the beta index. We adopt for the temperature profile $T(r)$
the polytropic form of
\begin{equation}
T(r) = T_0 \, [n(r)/n_0]^{\gamma - 1},
\label{eq:polytropic}
\end{equation}
where $T_0$ is the temperature at $r=0$, and $\gamma$ is the polytrope
index. The simulated profiles show reasonable agreement with the above
models. The best-fit values of $\beta$ and $\gamma$ are listed in Table
\ref{tab:all}.  The range of $\gamma$ is approximately 1.1-1.2.

In addition to their radial gradients, the gas density and temperature
have small-scale fluctuations. Figure \ref{fig:LNall} illustrates the
distributions of the gas density and temperature in each radial shell
normalized by their averaged quantities,$\A T \E$ and $\A n \E$,
respectively. Despite some variations among different shells, we find a
striking similarity in the overall shape of the distributions. They
approximately follow the log-normal distribution given by
\begin{eqnarray}
\PLN(\Dx) \, d \, \Dx =  
\frac{1}{\sqrt{2 \pi} \Sigx}
 \exp{\left[ \frac{-\left(\log{\Dx}+\Sigx^2/2 \right)^2}{2 \Sigx^2}
      \right]} \, \frac{d \Dx}{\Dx},
\label{eq:pdf_delta}
\end{eqnarray}
where $\Dx \equiv x/\A x \E$ and $x$ denotes $T$ or $n$ ($\DT \equiv
T/\A T \E$,$\Dn \equiv n/\A n \E$).  For simplicity, we neglect the
variations among different shells and fit the distribution for the whole
cluster within $r_{200}$ (solid line) by the above equation (dashed
line). The best-fit values of $\SigT$ and $\Sign$ are listed in Table
\ref{tab:all}.

The small-scale fluctuations mentioned above are not likely an artifact
of the SPH scheme. We have applied the similar analysis to the data of
grid-based simulations (D.Ryu, private communication) and obtained
essentially the same results. Thus the log-normal nature of the
fluctuations is physical, rather than numerical.

\subsection{Analytical Model}

Based on the distributions of gas density and temperature described in
\S 3.1, we develop an analytical model to describe the contributions
of the {\it radial profile} (RP) and the {\it local inhomogeneities}
(LI) to the bias in the spectroscopic  temperature.

To describe the emission-weighted and spectroscopic-like temperatures in
simple forms, let us define a quantity:
\begin{eqnarray}
A_\alpha &\equiv& \int n(\rv)^2 T(\rv)^\alpha \, d{\bf r}, \nonumber \\
&=& \int_0^R r^2 dr \, \int d\Omega \, n^2({\bf r}) T^\alpha({\bf r}), 
\label{eq:Aalpha}
\end{eqnarray}
where the second line is for the spherical coordinate and $R$ denotes
the maximum radius considered here (we adopt
$R=r_{\mathrm{200}}$). Using this quantity, we can write down $\Tsl$ via
equation (\ref{eq:def_tsl}) as
\begin{eqnarray}
\Tslm=A_{a-1/2}/A_{a-3/2}.
\label{eq:tsl_a}
\end{eqnarray}
When the temperature is higher than $\sim 3$ keV, the cooling function
is dominated by the thermal bremsstrahlung; $\Lambda \propto
\sqrt{T}$. We find that substitution $\sqrt{T}$ for $\Lambda (T)$ yields
only $\sim 1 \%$ difference for the simulated clusters.  In the present
model, we adopt for simplicity the thermal bremsstrahlung cooling
function, $\Lambda \propto \sqrt{T}$, and then equation
(\ref{eq:def_tew}) reduces to
\begin{eqnarray}
\Tewm=A_{3/2}/A_{1/2}.
\label{eq:tew_a}
\end{eqnarray}

 When evaluating equation (\ref{eq:Aalpha}), we replace the spatial
average with the ensemble average:
\begin{eqnarray}
\int d\Omega \, n^2({\bf r}) T^\alpha({\bf r}) 
&=& 
4 \pi  [\A n \E (r)]
 ^2 [\A T \E (r)]^\alpha \int \int
  \Dn^2 \DT^\alpha
    P(\Dn,\DT;r) d\Dn d\DT, 
\end{eqnarray}
where $P(\Dn,\DT; r)$ is a joint probability density function at $r$.
Assuming further that the temperature inhomogeneity is uncorrelated with
that of density, i.e. $P(\Dn,\DT;r) = P(\Dn;r)\,P(\DT;r)$, we obtain
\begin{eqnarray}
A_{\alpha} = 4\pi \int_0^R r^2 [\A n \E (r)]^2 
[\A T \E(r)]^\alpha dr \int_0^{\infty} 
 \Dn^2 P(\Dn;r)
 d\Dn \int_0^{\infty}  \DT^\alpha P(\DT;r) d\DT. 
\label{eq-aalpha}
\end{eqnarray}
For the log-normal distribution of the temperature and the density
fluctuations, the average quantities are expressed as
\begin{eqnarray}
\int_0^{\infty} \DT^\alpha P(\DT;r) d\DT &=& 
\exp{\left[ \frac{\alpha
						     (\alpha-1) 
					\SigT (r)^2}{2} \right]}, \\ 
\int_0^{\infty} \Dn^2 P(\Dn;r) d\Dn  &=&  \exp{[\Sign
					(r)^2]}.  \label{eq:ndl}
\end{eqnarray}
If $\SigT$ and $\Sign$ are independent of the radius ($\SigT (r) = \SigT$
, $\Sign(r) = \Sign $), equation (\ref{eq-aalpha}) reduces to
\begin{eqnarray}
A_\alpha &=& \exp{(\Sign ^2)} \, \exp{\left[\frac{\alpha(\alpha-1)}{2} \, \SigT  ^2
			\right]} \, \times 4 \pi  \int_0^R r^2 \,  \, [\A n \E (r)]^2 \, \, [\A T \E (r)]^\alpha d r. 
\label{eq:a_RPxLI} 
\end{eqnarray}

Using the above results, $\Tslm$ and $\Tewm$ are expressed as 
\begin{eqnarray}
\label{eq:tsl1}
\Tslm&=&A_{a-1/2}/A_{a-3/2} = \TslP \exp{ \left[ \left(a-\frac{3}{2} \right) \SigT^2 \right]}, \\
\Tewm&=&A_{3/2}/A_{1/2} = \TewP \exp{\left(\frac{1}{2} \SigT^2 \right) },
\label{eq:tew1}
\end{eqnarray}
where $\TslP$ and $\TewP$ are defined as 
\begin{equation}
\TslP \equiv  \frac{\int_0^R r^2 \,  \, [\A n \E (r)]^2 \, \, [\A T \E
 (r)]^{(a-1/2)} d r }{\int_0^R r^2 \,  \, [\A n \E (r)]^2 \, \, [\A T \E (r)]^{(a-3/2)} d r },
\label{eq:tslp}
\end{equation}
\begin{equation}
\TewP \equiv = \frac{\int_0^R r^2 \, \, [\A n \E (r)]^2 \, \, [\A T \E
 (r)]^{3/2} d r }{\int_0^R r^2 \,  \, [\A n \E (r)]^2 \, \,  [\A T \E (r)]^{1/2} d r }.
\label{eq:tewp}
\end{equation}
As expected, $\Tslm$ and $\Tewm$ reduce to $\TslP$ and $\TewP$ in the
absence of local inhomogeneities ($\SigT = 0$). Note that equations
(\ref{eq:tsl1}) and (\ref{eq:tew1}) are independent of $\Sign$. 
This holds true as long as the density distribution $P(\Dn)$ is
independent of $r$.

The ratio of $\Tslm$ and $\Tewm$ is now written as 
\begin{eqnarray}
\Kapmo \equiv \Tslm/\Tewm = \KP \KL, 
\label{eq:kapall}
\end{eqnarray}
where $\KP$ and $\KL$ denote the bias due to the radial profile and the
local inhomogeneities,
\begin{equation}
\KP \equiv \TslP/\TewP,
\label{eq:kapp}
\end{equation}
and 
\begin{eqnarray}
\KL \equiv  \exp{\left[(a - 2) \, \SigT^2 \right]},
\label{eq:kl}
\end{eqnarray}
respectively. Figure \ref{fig:KL} shows $\KL$ for a fiducial value of
$a=0.75$ as a function of $\SigT$. The range of $\SigT$ of simulated
clusters, $0.1<\SigT<0.3$, (Table \ref{tab:all}) corresponds to
$0.99>\KL>0.89$.

In the case of the beta model density profile (eq.[\ref{eq:beta-model}])
and the polytropic temperature profile (eq.[\ref{eq:polytropic}]),
$\TslP$ and $\TewP$ are expressed as
\begin{eqnarray}
\KP &=& \TslP/\TewP, \\
\TslP &=& \frac{\,_2 F_1 (3/2,3 \beta [1 + (\gamma-1)(a-1/2)/2
 ] ; 5/2 ; -R^2/r_c^2 )}{\,_2 F_1 (3/2,3 \beta
 [1 + (\gamma-1)(a-3/2)/2] ; 5/2 ; -R^2/r_c^2 )} T_0, \\
\TewP &=& \frac{\,_2 F_1 (3/2,3 \beta [1 + 3(\gamma-1)/4 ]
 ; 5/2 ; -R^2/r_c^2 )}{\,_2 F_1 (3/2,3 \beta [1
 + (\gamma-1)/4] ; 5/2 ; -R^2/r_c^2 )} T_0, 
\label{eq:tewp_poly}
\end{eqnarray}
where $\,_2 F_1 ( \alpha,\beta; \gamma; \zeta)$ is the hyper geometric
function.

Figure \ref{fig:kapparp} shows $\KP$ as a function of $\beta$ for
various choices of $\gamma$ and $r_c/r_{200}$. Given that a number of
observed clusters exhibit a cool core, we also plot the case with the
temperature profile of the form \citep{allen01,kaastra04}:
\begin{equation}
T(r) = T_l + (T_h - T_l) \, \frac{(r/r_c)^\mu}{1 + (r/r_c)^\mu}, 
\label{eq:cool}
\end{equation}
with $(T_h - T_l)/T_l = 1.5$ and $\mu=2$. For the range of parameters
considered here, $\KP$ exceeds 0.9. This implies that the bias in the
spectroscopic temperature is not fully accounted for by the global
temperature and density gradients alone; local inhomogeneities should
also make an important contribution to the bias.

\section{Comparison with Simulated Clusters}

We now examine the extent to which the analytical model described in the
previous section explains the bias in the spectroscopic temperature.
The departure in the radial density and temperature distributions from
the beta model and the polytropic model results in up to 7 \% errors in
the values of $\TewP$ and $\TslP$. Since our model can be applied to
arbitrary $\A n \E (r)$ and $\A T \E (r)$, we hereafter use for these
quantities the radially averaged values calculated directly from the
simulation data.  We combine them with $\SigT$ in Table \ref{tab:all} to
obtain $\Tslm$ (eq.[\ref{eq:tsl1}]) and $\Tewm$ (eq.[\ref{eq:tew1}]).
 
Figure \ref{fig:tslf} compares $\Tslm $ and $\Tewm $ against $\Tslsm$
and $\Tewsm$ (eq.[\ref{eq:tslmesh}] and eq.[\ref{eq:tmesh}]),
respectively.  For all clusters except ``Perseus'', the model reproduces
within 10 percent accuracy the temperatures averaged over all the mesh
points of the simulated clusters.

 Given the above agreement, we further plot $\Kapmo$ against $\Kapsim$
in Figure \ref{fig:kapf}. The difference between $\Kapsim$ and $\Kapmo$
is kept within $\sim 10$ \% in all the cases. Considering the simplicity
of our current model, the agreement is remarkable.  In all the clusters,
both $\KP$ and $\KL$ are greater than $\Kapsim$, indicating that their
combination is in fact responsible for the major part of the bias in the
spectroscopic temperature.

\begin{figure}
\plotone{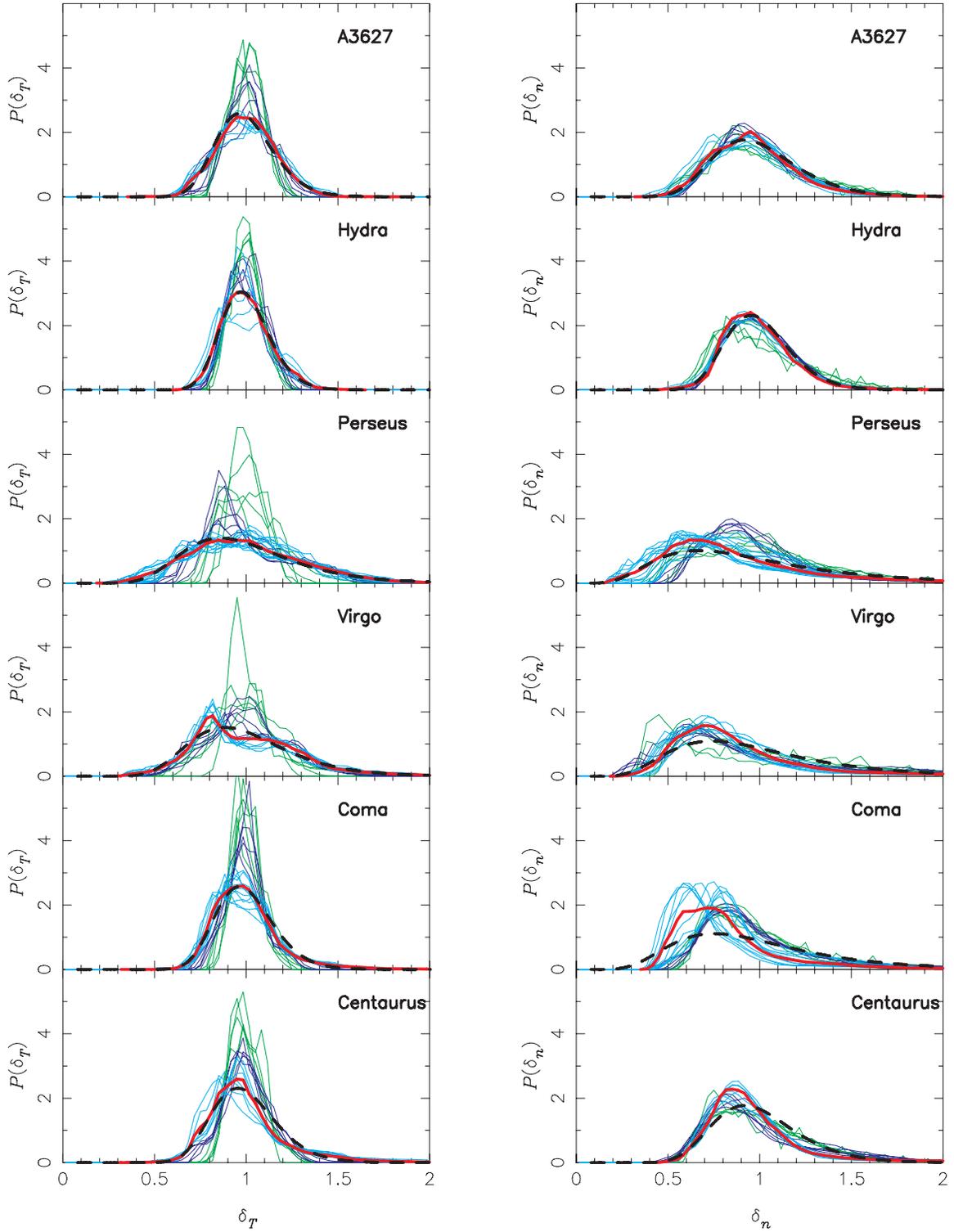}
  \caption{The distribution of $\DT  \equiv T/\A T \E$ and
 $\Dn \equiv n/\A n \E$. Thick solid lines 
 present the distribution throughout the mesh points of
 within $r=r_{200}$. Dashed lines are
 fitting lines of the log-normal distribution. Thin solid lines are the
 distribution of the shells each  $67 h^{-1}~ \mathrm{kpc}$ distance
 from the center. Each color indicates different radial
 interval: $r < 335 h^{-1}~ \mathrm{kpc}$ (green), $335 h^{-1}~
 \mathrm{kpc}< r < 670 h^{-1}~ \mathrm{kpc}$ (blue), and $r > 675
 h^{-1}~ \mathrm{kpc}$ (cyan). } 
\label{fig:LNall}
\end{figure}

\begin{figure}[ht]
\plotone{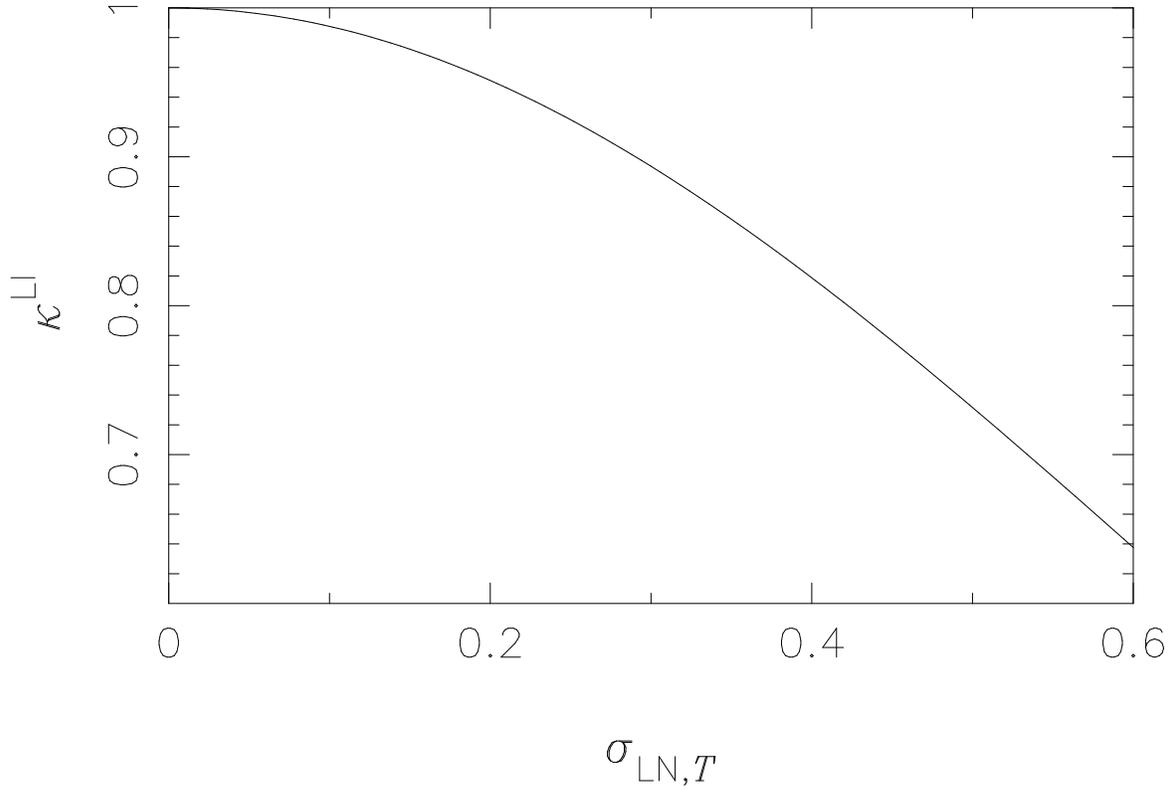}
  \caption{The shape of $\KL (\SigT)$ adopting $a=0.75$. 
  In the case of
 $\SigT = 0.1$ and $0.3$, $\KL \sim 0.99$ and $0.89$, respectively (See
 Table \ref{tab:all}). } 
 \label{fig:KL}
\end{figure}

\begin{figure}[htbp]
 \plotone{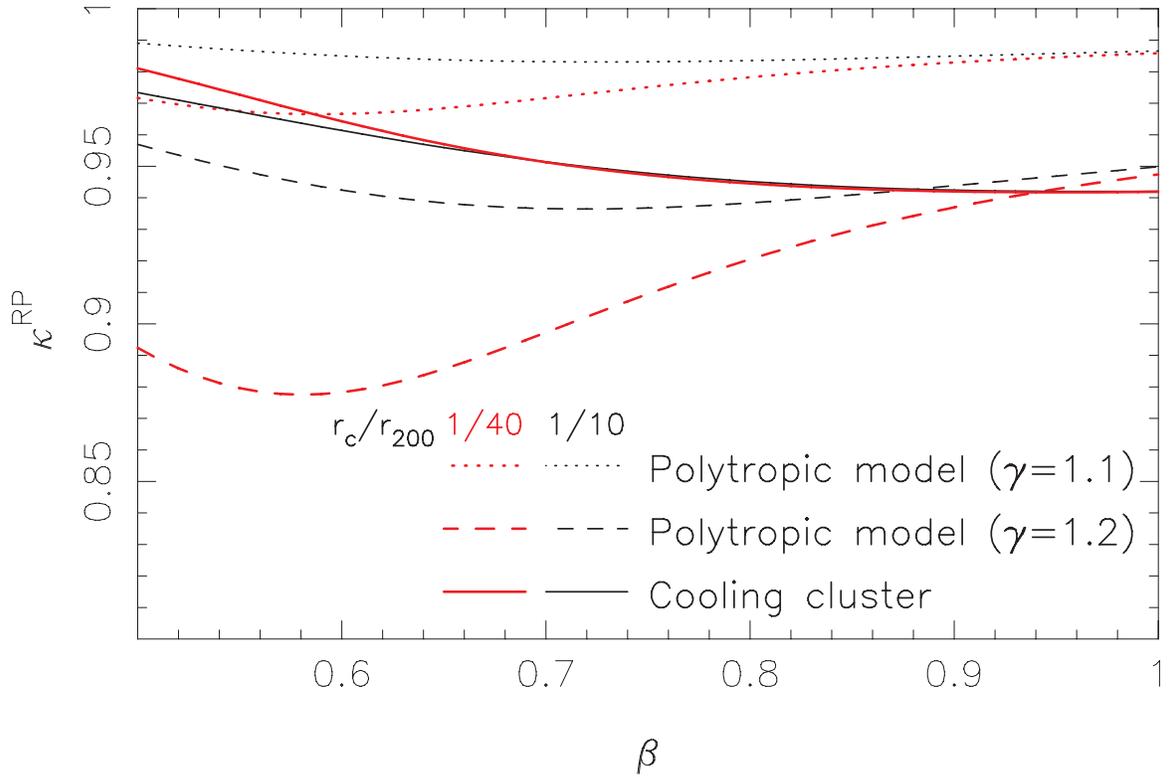}
  \caption{The bias due to the radial profile $\KP$ (eq.[\ref{eq:kapp}]) 
 assuming the beta model and two temperature 
 models. We assume that the density profile is given by the
 beta-model. We consider three temperature models: the polytropic model
 (dotted line provides $\gamma = 1.1$, dashed line $\gamma=1.2$) and the
 cooling cluster model (solid line). We assume the two case of
 $r_c/r_{200}$. One is $r_c/r_{200}=1/10$ (black line). Another is $r_c/r_{200}=1/40$ (red line). }
\label{fig:kapparp}
\end{figure}

\begin{figure}[htbp]
\plotone{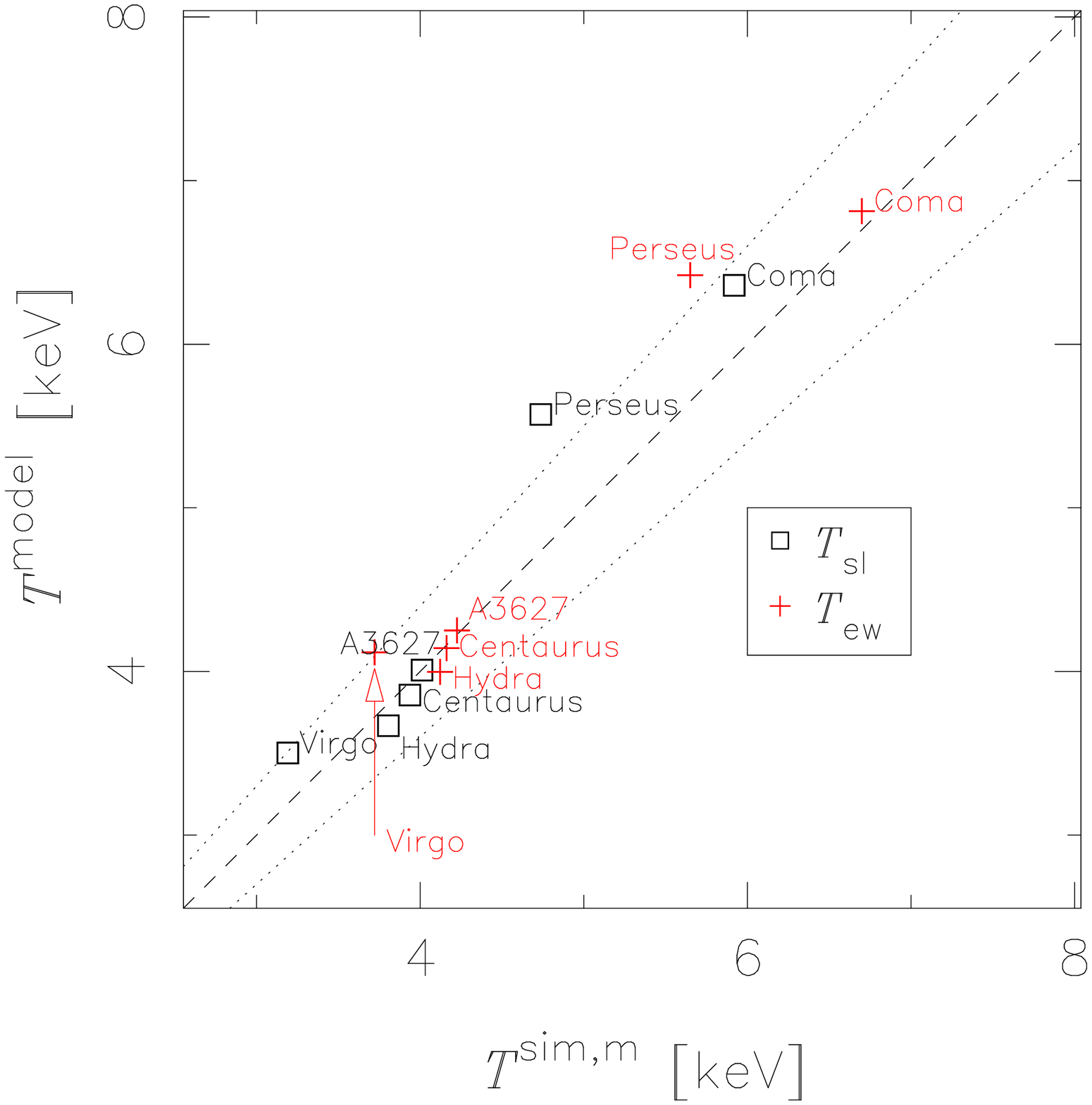}
  \caption{The emission weighted and spectroscopic like temperature
 provided our model and the simulation. Dashed line shows $\Tewsm = \Tewm$ or $\Tslsm = \Tslm$.Dotted lines
 show $\Tewsm/\Tewm - 1 = \pm 0.1$ or $\Tslsm/\Tslm - 1 = \pm 0.1$. In
 all clusters except ``Perseus'', the temperatures of the
 model reproduce  
 that of the simulation within 10 percent.}
\label{fig:tslf}
\end{figure}

\begin{figure}[h]
\plotone{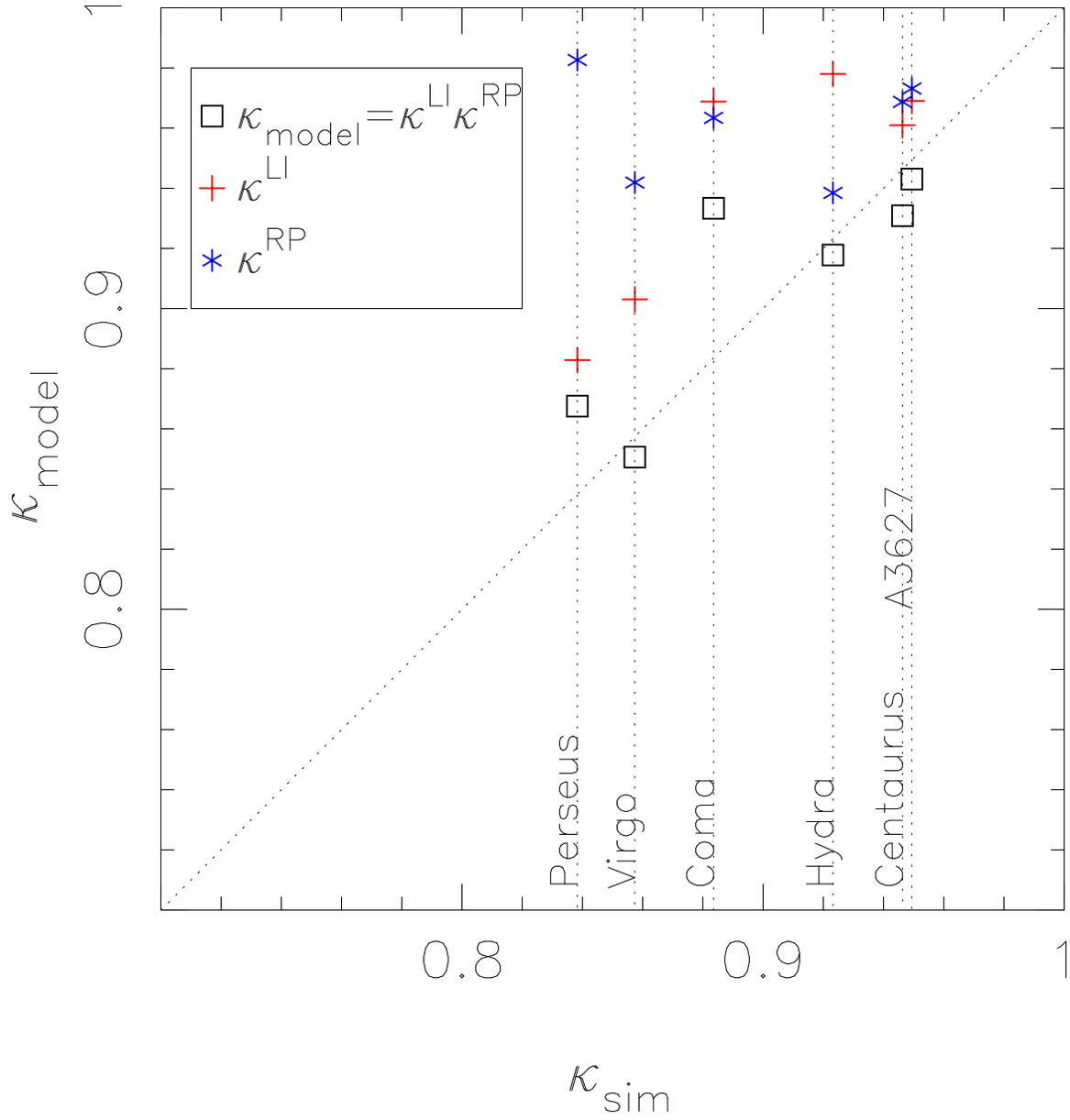}
  \caption{The bias factor $\kappa$ provided our
 model and the simulation. 
 Squares show $\Kapmo$. Asterisks and crosses show $\KP$
 and $\KL$ which are calculated from our model.
In all cases, $\Kapmo$ is kept within $\sim 10$ \%.
}
\label{fig:kapf}
\end{figure}

\begin{figure}[h]
\plottwo{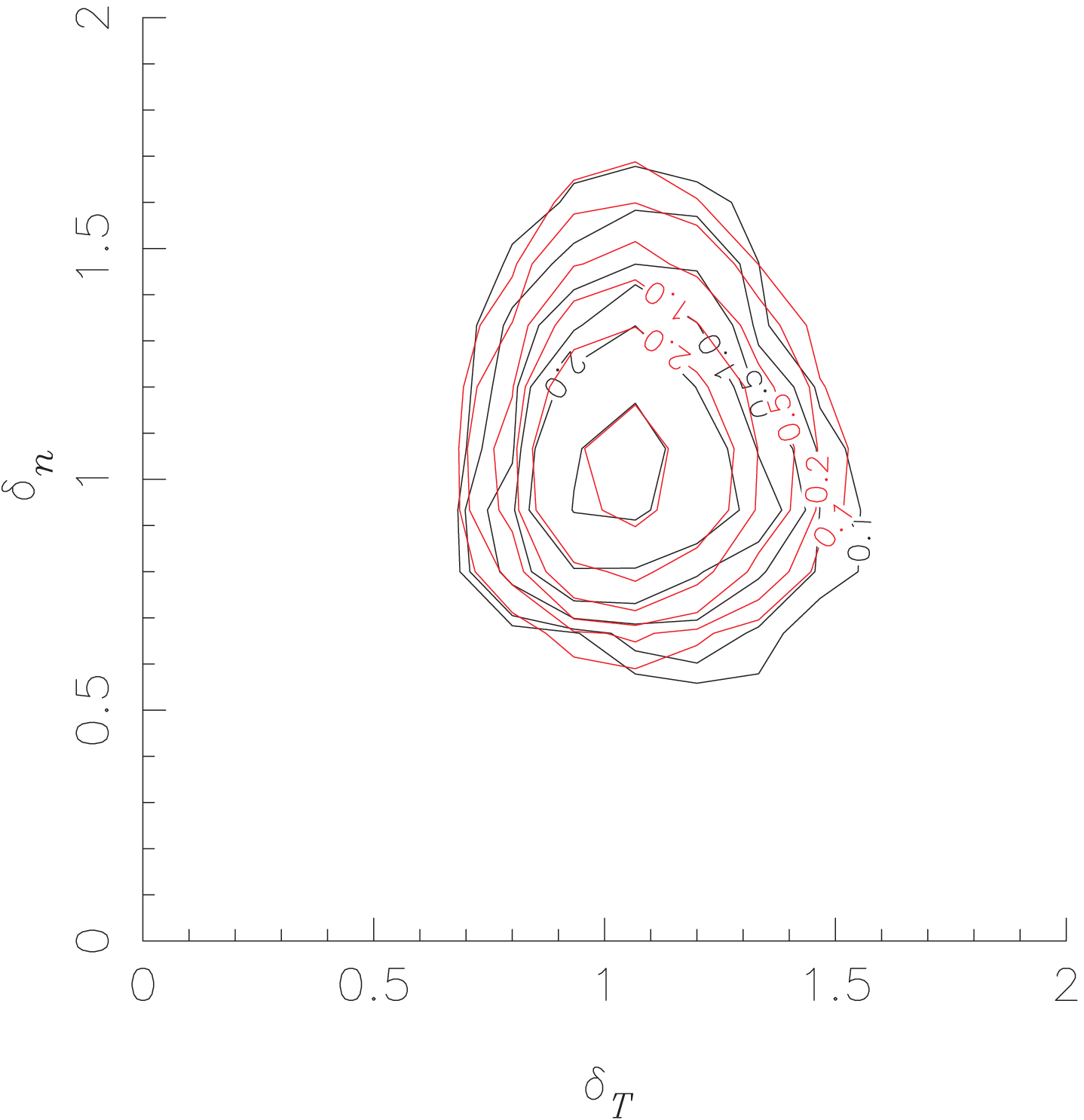}{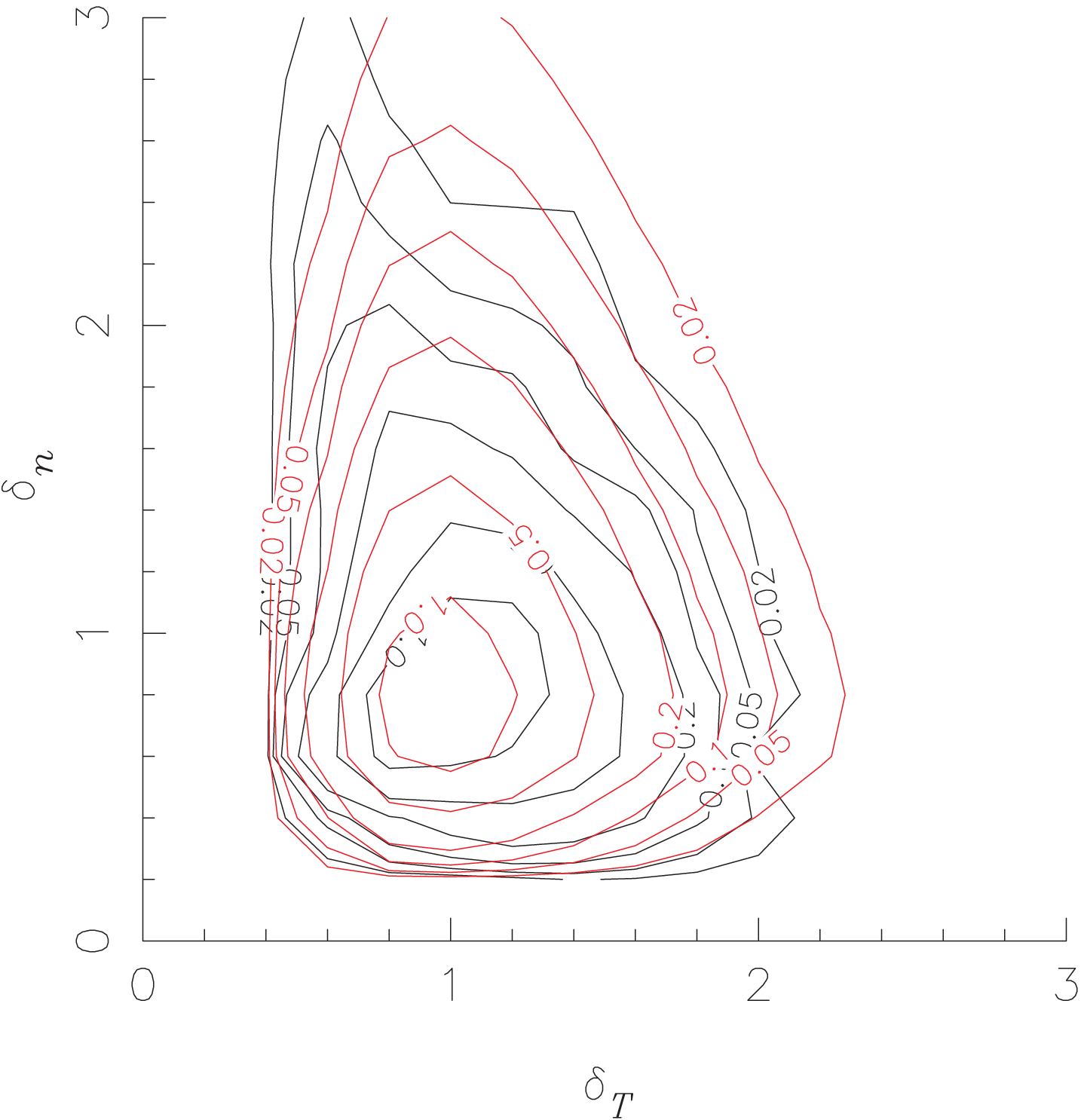}
  \caption{The contour map of the joint probability P($\DT$,$\Dn$) within
 $r_{200}$ (black contour). Red contour indicates the
 theoretical line of P($\DT$)P($\Dn$). Left panel shows the case of ``Hydra'',Right panel ''Perseus''.} 
\label{fig:deltat-deltarho}
\end{figure}

In \S 3.2, we assumed that $n$ and $T$ are uncorrelated, i.e.,
$P(\Dn,\DT)=P(\Dn)P(\DT)$. We examine this assumption in more detail. We
pick up two clusters, ``Hydra'' and ``Perseus'', which show the best and
the worst agreement, respectively, between $\Kapmo$ and $\Kapsim$.
Figure \ref{fig:deltat-deltarho} shows the contours of the joint
distribution of $\DT = T({\bf r })/\A T \E (r)$ and $\Dn = n({\bf r
})/\A n \E (r)$ for all the mesh points within $r_{200}$ for these
clusters, together with that expected from the model assuming
$P(\Dn,\DT) = \PLN(\Dn) \PLN(\DT)$. For the log-normal distributions,
$\PLN(\Dn)$ and $\PLN(\DT)$, we have used the fits shown as the dashed
line in Figure \ref{fig:LNall}. The joint distribution agrees well with
the model for both cases, while the deviation is somewhat larger in
``Perseus'' for which the model gives poorer fits to the underlying
temperature and density distribution in Figure \ref{fig:LNall}.  If the
cluster is spherically symmetric and the ICM is in hydrostatic
equilibrium, we expect that $n$ are correlated with $T$ as $\Dn \DT =
1$. We do not find such correlations in Figure
\ref{fig:deltat-deltarho}.

 Although the effects of the correlation is hard to model in general, we
can show that it does not change the value of $\Kapmo\equiv\Tslm/\Tewm$
as long as the joint probability density function follows the bivariate
log-normal distribution:
\begin{eqnarray}
\PBLN(\Dn,\DT) \, d\Dn \, d\DT = \frac{(1 - \rho'^2)^{-1/2}}{2
 \pi \Sign \SigT} \exp{\left[-\frac{A^2 - 2 \rho' A B + B^2}{2
				   (1-\rho'^2)}\right]} 
 \, \frac{d\Dn}{\Dn} \, \frac{d\DT}{\DT},   
\end{eqnarray}
where $\rho' \equiv
\log{[\rho(\exp{\Sign^2}-1)^{1/2}(\exp{\SigT^2}-1)^{1/2} +1]}/\SigT
\Sign $, ~~$A \equiv \log{(\Dn)} + \Sign^2/2$, ~~$B \equiv \log{(\DT)} +
\SigT^2/2 $, and $\rho$ is the correlation coefficient between $n$ and
$T$. Adopting $\rho=0$ yields $\PBLN(\Dn,\DT)=\PLN(\Dn)
\,\PLN(\DT)$. The marginal probability density function of density $\int
d\DT \PBLN(\Dn,\DT) $ and that of temperature $\int d\Dn \PBLN(\Dn,\DT)
$ are equal to $\PLN(\Dn)$ and $\PLN(\DT)$, respectively. Using
$\PBLN(\Dn,\DT)$, we obtain $\Tslm$, $\Tewm$ as
\begin{eqnarray}
\Tslm&=&\TslP \exp{ \left[ \left(a-\frac{3}{2} \right) \SigT^2 + 2 \rho' \SigT \Sign \right]} \\
\Tewm&=&\TewP \exp{\left(\frac{1}{2} \SigT^2 + 2 \rho' \SigT \Sign\right) }. 
\label{eq:tewc}
\end{eqnarray}
Although both $\Tslm$ and $\Tewm$ increase with the correlation
coefficient, $\Kapmo\equiv\Tslm/\Tewm$ remains the same as that given by
equation (\ref{eq:kl}).

\section{Summary and Conclusions}

We have explored the origin of the bias in the spectroscopic temperature
of simulated galaxy clusters discovered by \citet{mazzotta04}.  Using
the independent simulations data, we have constructed mock spectra of
clusters, and confirmed their results; the spectroscopic temperature is
systematically lower than the emission-weighted temperature by 10-20\%
and that the spectroscopic-like temperature defined by equation
(\ref{eq:def_tsl}) approximates the spectroscopic temperature to better
than $\sim$ 6\%.  In doing so, we have found that the multi-phase nature
of the intra-cluster medium is ascribed to the two major contributions,
the radial density and temperature gradients and the local
inhomogeneities around the profiles. More importantly, we have shown for
the first time that the probability distribution functions of the local
inhomogeneities approximately follow the log-normal distribution. Based
on a simple analytical model, we have exhibited that not only the radial
profiles but also the local inhomogeneities are largely responsible for
the above mentioned bias of cluster temperatures.

We would like to note that the log-normal probability distribution
functions for density fields show up in a variety of
astrophysical/cosmological problems
\citep[e.g.,][]{Hubble1934,CJ1991,WN2001,Kayo2001,Taruya2002}.  While it
is not clear if they share any simple physical principle behind, it is
interesting to attempt to look for the possible underlying connection.

In this paper, we have focused on the difference between spectroscopic
(or spectroscopic-like) and emission-weighted temperatures, which has
the closest relevance to the X-ray spectral analysis. Another useful
quantity is the mass-weighted temperature defined by
\begin{equation}
\Tmw \equiv \frac{\int n T dV}{\int n dV}.
\label{eq:def_tmw}
\end{equation}
This is related to the cluster mass more directly
\citep[e.g.,][]{mathiesen01,nagai06}.  The mass-weighted temperature is
highly sensitive to the radial density and temperature profiles, while
it is little affected by the local inhomogeneities.  Though challenging,
it will be ``observable'' either by high-resolution X-ray spectroscopic
observations \citep{nagai06} or by a combination of the lower resolution
X-ray spectroscopy and the Sunyaev-Zel'dovich imaging observations
\citep{Komatsu99,Komatsu01,Kitayama04}.  We
will discuss usefulness of this quantity and the implications for the
future observations in the next paper.

\vspace*{0.5cm}
\noindent

We thank Noriko Yamasaki and Kazuhisa Mitsuda for useful discussions,
Dongsu Ryu for providing the data of grid-based simulations, and Luca
Tornatore for providing the self-consistent feedback scheme handling the
metal production and metal cooling used for the hydrodynamical
simulation. We also thank the referee, Stefano Borgani, for several
pertinent comments on the earlier manuscript. The hydrodynamical
simulation has been performed using computer facilities at the
University of Tokyo supported by the Special Coordination Fund for
Promoting Science and Technology, Ministry of Education, Culture, Sport,
Science and Technology.  This research was partly supported by
Grant-in-Aid for Scientific Research of Japan Society for Promotion of
Science (Nos.\ 14102004, 18740112, 15740157, 16340053). 
Support for this work was provided partially by NASA through Chandra
Postdoctoral Fellowship grant number PF6-70042 awarded by the Chandra
X-ray Center, which is operated by the Smithsonian Astrophysical
Observatory for NASA under contract NAS8-03060.

\newpage

\begin{table}
\caption{Properties of the six simulated clusters and observed clusters.}
\label{tab:all}
 \begin{center}
  \begin{tabular}{lccccccc}
   \hline\hline
   & & & Simulation & & & \\
   \hline
   \multicolumn{1}{c}{} & A3627 & Hydra & Perseus & Virgo &
   Coma & Centaurus & \\
   \hline
$M_{200} [10^{14} h^{-1} M_{\odot}]$ & 2.2 & 1.8 & 6.7 & 3.1 & 4.3 & 2.5 &\\ 
$r_{200}$ [$h^{-1}$ Mpc]& 1.1 & 1.0 & 1.6 & 1.2 & 1.4 & 1.1 &\\ 
$\beta$ & 0.69 & 0.72 & 0.65 & 0.59 & 0.73 & 0.73 &\\ 
$\gamma$ & 1.21 & 1.26 & 1.09 & 1.12 & 1.19 & 1.14 &\\ 
$\Tew$ [keV]& 4.2 & 4.1 & 5.7 & 3.7 & 6.7 & 4.2 &\\ 
$\Tsl$ [keV]& 4.0 & 3.8 & 4.7 & 3.2 & 5.9 & 3.9 &\\ 
$\Tspec$ (IDEAL) [keV] (0.5-10.0 keV)& 4.1 & 3.9 & 5.0 & 3.3 & 6.1 & 3.9 &\\
$\Tspec$ (ACIS) [keV] (0.5-10.0 keV)& 4.1 & 3.9 & 5.0 & 3.3 & 6.0 & 4.0 &\\ 
$\Tspec$ (MOS) [keV] (0.5-10.0 keV)& 4.0 & 3.8 & 5.0 & 3.3 & 6.0 & 4.0 &\\ 
$\Tspec$ (IDEAL) [keV] (0.1-2.4 keV)& 4.0 & 3.7 & 4.7 & 3.1 & 5.9 &  3.9 &\\
$\Tspec$ (IDEAL) [keV] (2.0-10.0 keV)& 4.1 & 4.0 & 5.4 & 3.6 & 6.5 &  4.0 &\\
$\Kapsim$ (mesh-wise)& 0.95 & 0.92 & 0.84 & 0.86 & 0.88 & 0.95 &\\ 
$\Kapsim$ (particle-wise)& 0.88 & 0.89 & 0.70 & 0.75 & 0.84 & 0.86 &\\
$\KP$ & 0.97 & 0.94 & 0.98 & 0.94 & 0.96 & 0.97 & \\
$\SigT$ & 0.159 & 0.180 & 0.316 & 0.286 & 0.159 & 0.178 &\\
$\Sign$ &  0.240 & 0.180 & 0.518 & 0.446 & 0.434 & 0.239 &\\
   \hline\hline
  \end{tabular}
  \begin{tabular}{lccccccc}
   & & & Observation & & & &  \\  
   \hline
   \multicolumn{1}{c}{} & A3627 & Hydra & Perseus & Virgo &
   Coma & Centaurus & Ref\\
\hline
$M_{200} [10^{14} h^{-1} M_{\odot}]$ & $^*4.6^{+0.81}_{-0.58}$ & $1.90^{+0.38}_{-0.33}$ & $9.08^{+2.13}_{-1.52}$ & $2.04^{+0.28}_{-0.21}$ & $4.97^{+0.68}_{-0.57}$ & $6.97^{+1.22}_{-1.25}$ & 1,$^*$2\\ 
$r_{200}$ [$h^{-1}$ Mpc]&  $^*$1.26 & 1.22 & 2.05 & 1.26 & 1.64 & 0.89 &   1,$^*$2\\
$\Tspec$ [keV]& $5.62^{+0.12}_{-0.11}$& $3.82^{+0.20}_{-0.17}$ &
   $6.42^{+0.06}_{-0.06}$ & $^\dagger$$2.5^{+0.04}_{-0.05}$ & $8.07^{+0.29}_{-0.27}$ & $3.69^{+0.05}_{-0.04}$ & 3,$^\dagger$4\\
$N_{\mathrm{H}}$ [$10^{20} \mathrm{cm^{-2}}$]& 21.7& 4.79 & 13.9& 2.58 & 0.93 &
   8.1 & 5\\
\hline 
     \multicolumn{8}{@{}l@{}}{\hbox to 0pt{\parbox{200mm}{\footnotesize
\par\noindent
References.--(1)\cite{girardi1998};(2)\cite{reiprich2002};(3)\cite{ikebe2002}
\par\noindent
(4)\cite{shibata2001};(5)\cite{dickey90} 
   }\hss}}
  \end{tabular}
 \end{center} 
\end{table}

\end{document}